\documentclass[,final]{aipproc}

\layoutstyle{6x9}

\usepackage{amssymb}
\usepackage{amsmath}
\usepackage{graphicx}
\usepackage{multicol}
\usepackage{floatflt}
\begin{document}

\title{Recent measurement of $\Delta G/G$ at COMPASS}



\author{C. Bernet, on behalf of the COMPASS collaboration}{address=colin.bernet@cern.ch}

\begin{abstract}
We present a preliminary measurement of the gluon polarization $\Delta G/G$ 
in the nucleon, based on the spin asymmetry of quasi-real photoproduction
events for which a pair of large transverse momentum hadrons is produced. The data
were obtained by the COMPASS experiment at CERN using a 160 GeV 
polarized muon beam scattered on a large polarized $^6$LiD target.
The preliminary helicity asymmetry for the selected events is $A_{\parallel}/D = 0.002 \pm 0.019({\rm stat.}) \pm 0.003({\rm
exp.syst.})$. From this value, a leading order analysis based on the
model of PYTHIA leads to the gluon polarization in the nucleon
$\Delta G/G(x_g=0.095,\mu^2=3\ {\rm GeV}^2)=0.024 \pm 0.089({\rm stat.})
\pm 0.057({\rm syst.})$. This value is consistent with parameterizations obtained from QCD fits to the $g_1$ data, with a first moment $\Delta G \equiv \int_0^1 \Delta G(x) dx \lesssim 1$, at the same scale.
\end{abstract}

\maketitle

The decomposition of the nucleon spin in the contributions from
its constituents has been a central topic of investigation in
polarized lepton-nucleon scattering in the last 20 years. The
EMC measurement of the proton spin structure~\cite{EMC:1988} has shown
that only 20 to 30~\% of the proton spin could be attributed to
the total quark spin $\Delta \Sigma$, in contrast to the 60~\%
expected in the naive quark-parton model. 
In inclusive lepton-nucleon scattering
the contribution of the gluon spin $\Delta G$ to the nucleon spin
can only be measured indirectly
by studying the $Q^2$ dependence of the polarized spin-structure functions
in QCD. Fits at next-to-leading order (NLO) provide evaluations of $\Delta G$
which are of the order of 0.50. The precision of these fits is strongly
limited by the small $Q^2$ range covered by the data at any value of $x$,
a situation resulting from the lack of a polarized lepton-nucleon collider.
In addition, the shape of $\Delta G(x)$ at a fixed $Q^2$ used as reference, has to be
provided as an input parameterization. It varies considerably between the
different analyses and is only poorly constrained by the results of the fits. A direct measurement of the gluon polarization $\Delta G(x)/G(x)$ can be obtained from the helicity asymmetry of the photon-gluon fusion (PGF, $\gamma^* g \rightarrow q \bar q$) cross-section, which constitutes an important part
of the experimental program of COMPASS

The COMPASS experiment \cite{Mallot:2004gk} is located at the M2 beam line of the CERN SPS, which provides a 160 GeV $\mu^+$ beam, with a natural polarization of $-76\pm5\%$. The target consists in an upstream and a downstream cell, longitudinally polarized in opposite directions. Typical target polarizations of $50.0 \pm 2.5 \%$ are obtained. The forward spectrometer is divided in two stages allowing the reconstruction of the scattered muon and of the produced hadrons in broad momentum and angular ranges. The trigger system provides efficient tagging down to $Q^2=0.002$~GeV$^2$.
%
%
%

The present analysis focuses on the data collected during the 2002 and 2003 runs. We only consider quasi-real photoproduction events ($Q^2<1$ GeV$^2$), in which at least two charged hadrons are associated to the primary vertex in addition to the incident and the scattered muons. The fraction of PGF is enhanced by asking the two leading hadrons to have a large transverse momentum: $p_T^{h1}>0.7$ GeV, $p_T^{h2}>0.7$ GeV and $(p_T^{h1})^2 + (p_T^{h2})^2>2.5$ GeV$^2$. In total, around 350,000 events are selected. For this {\em high $p_T$} sample, the measured helicity asymmetry (defined as in~\cite{Ageev:2005g1}) is 
\begin{equation}
\frac{A_\parallel}{D} = 0.002 \pm 0.019 (stat) \pm 0.003 (exp. syst),
\label{eq:measaparodall}
\end{equation}
where the quoted systematic error accounts for the false asymmetries related to the apparatus. Other sources of systematic errors, including the error on the beam and target polarizations, are only a few percents of the (small) measured asymmetry, and are therefore negligible. The 2004 data are currently under analysis, and represent the same amount of data as 2002 and 2003 altogether.

%
%
%
A Monte-Carlo simulation is needed to extract the gluon polarization from the high $p_T$ asymmetry. The selected  sample of high $p_T$ events covers the transition region ranging from photoproduction ($Q^2 \approx 0$) up to DIS ($Q^2 \approx 1$ GeV$^2$). For this reason, we chose PYTHIA as an event generator because it provides a model for the lepton-nucleon interactions~\cite{Friberg:2000ra} at low $Q^2$. Two different kinds of processes are generated. In the so-called {\em direct processes}, the virtual photon takes part in the hard partonic interaction. In the {\em resolved processes}, it fluctuates to a hadronic state, from which a parton is extracted. This parton then interacts with a parton from the nucleon. At $Q^2<1$, the resolved processes constitute half of the high $p_T$ sample. Their contribution falls to about 10\% for $Q^2>1$ GeV$^2$ and becomes negligible for $Q^2 > 2$~GeV$^2$. Note that the analysis requires the factorization between the hard and soft parts of the reaction, hence the presence of a hard scale $\mu^2$. As $Q^2<1$ GeV$^2$, the scale is provided by the transverse momentum of the partons involved in the reaction. Events for which no hard scale can be found are classified as {\em low $p_T$}.

The generated events are tracked through a GEANT description of the COMPASS spectrometer, and processed using the same reconstruction program as for real data. Then, the Monte-Carlo sample of high $p_T$ events is selected through the same cuts. 

\begin{figure}[b!]
\centering
\includegraphics[width=0.32\textwidth]{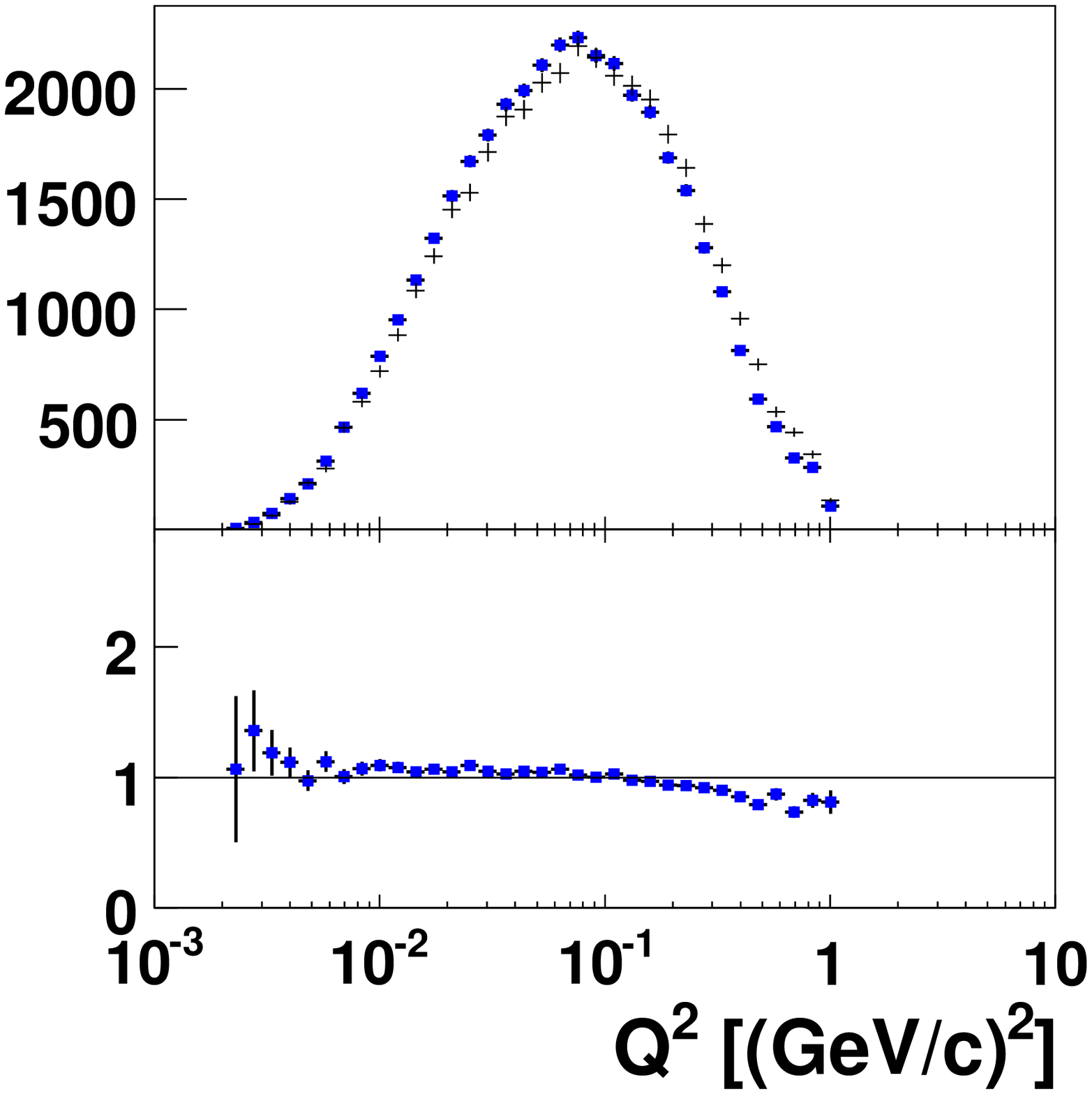}
\includegraphics[width=0.32\textwidth]{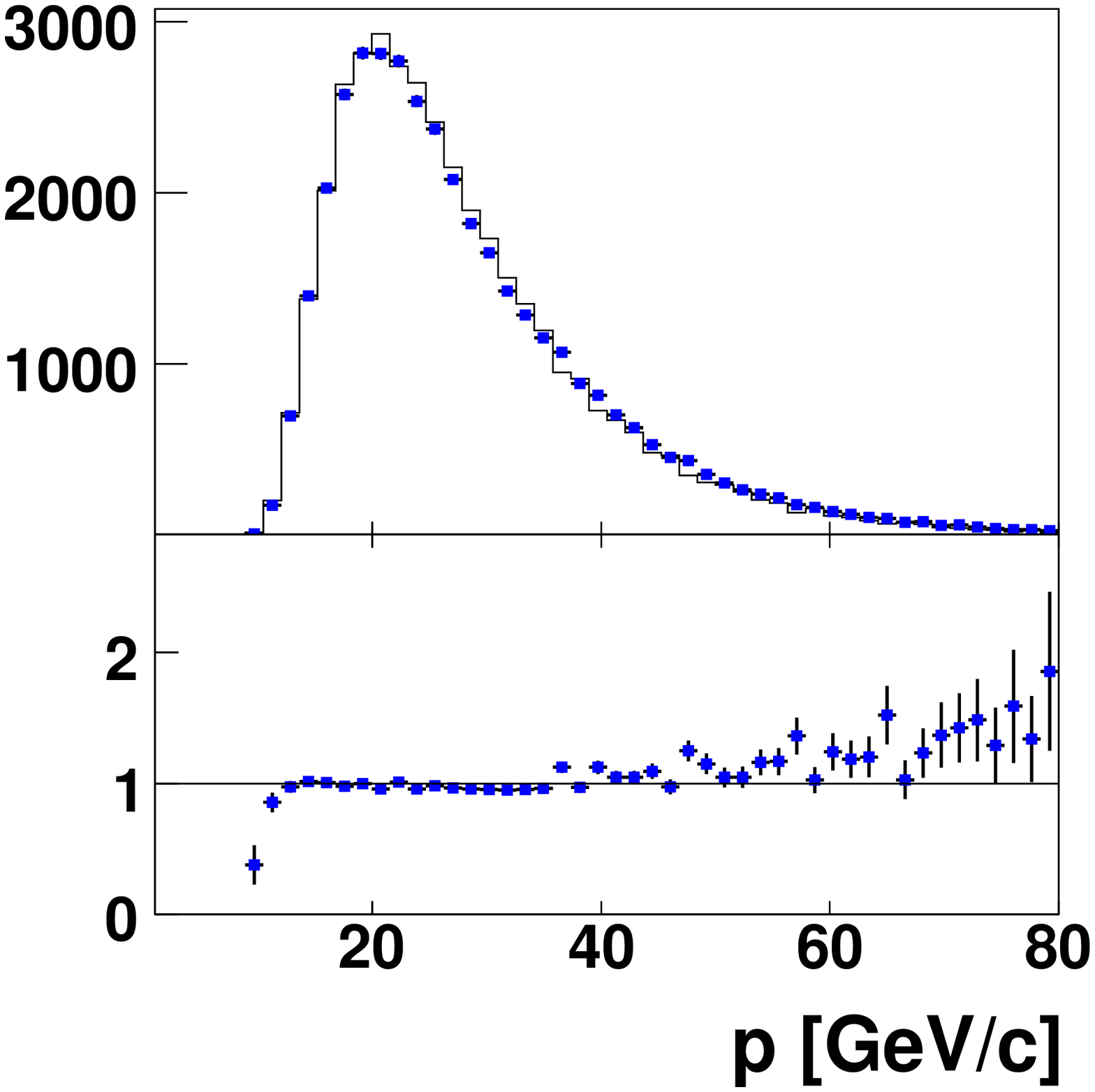}
\includegraphics[width=0.32\textwidth]{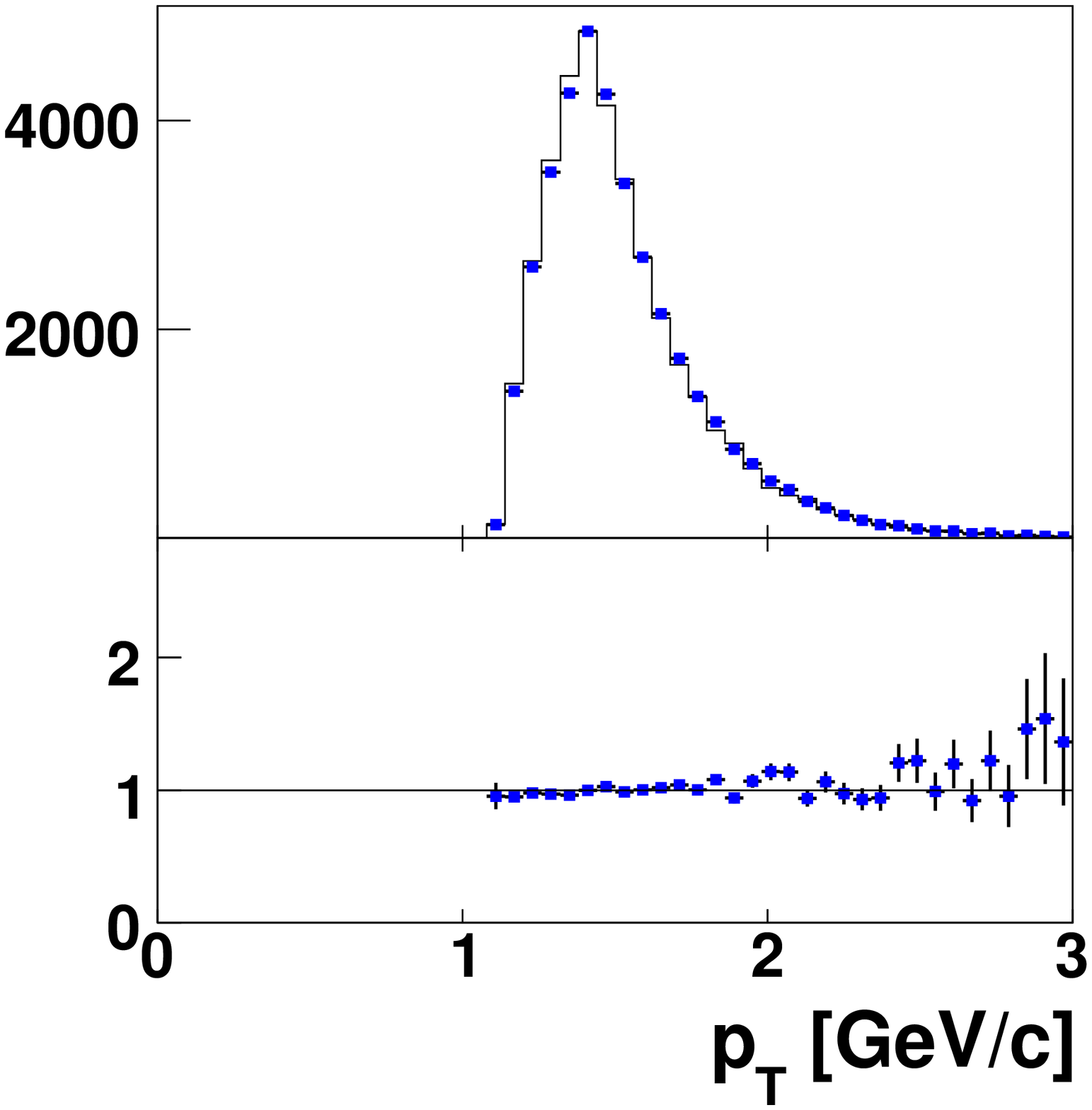}
\caption{The upper part of these plots show a comparison between the simulated (histogram) and real data (points) samples of high $p_T$ events, normalized to the number of events. The lower part shows the corresponding data/simulation ratio.
$p$ ($p_T$) is the total (transverse) momentum of the leading hadron. A similar agreement is obtained for the next-to-leading hadron. }
\label{fig:mcrealcomp}
\end{figure}

\begin{figure}
\centering
\includegraphics[width=0.8\textwidth]{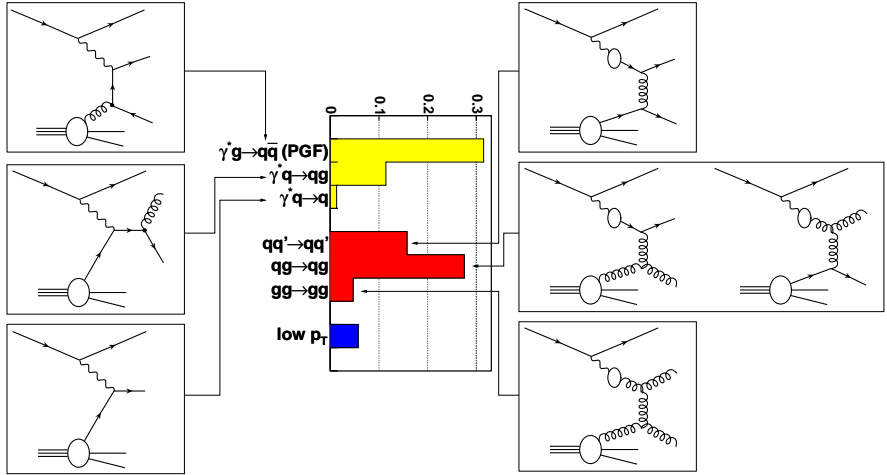}
\caption{Contribution of each PYTHIA subprocess to the Monte-Carlo sample of high $p_T$ events. On the left: direct processes (photon-gluon fusion, QCD Compton, and leading process); on the right: resolved processes; note that $q \bar q \rightarrow q' \bar q'$, $q \bar q \rightarrow gg$ and $gg \rightarrow q \bar q$ are neglected, as they altogether represent $0.6\%$ of the sample.}
\label{fig:rsubproc}
\end{figure}

Only one parameter of PYTHIA had to be changed to reach a good agreement with the data: the width of the Gaussian distribution of intrinsic transverse momentum of partons within the resolved virtual photon (PARP(99)) was decreased from 1~GeV/c to 0.5~GeV/c. Fig.~\ref{fig:mcrealcomp} presents a comparison between the simulated and real data samples of high $p_T$ events. Fig.~\ref{fig:rsubproc} shows how the Monte-Carlo sample of high $p_T$ events divides into the various PYTHIA subprocesses. 
%
%
%
%
The high $p_T$ asymmetry can be written in terms of the contributions of the different processes:
\begin{align}
  \nonumber \frac{A_\parallel}{D} = R_{PGF} \left\langle \hat a_{LL}^{PGF} / D \right\rangle \frac{\Delta G}{G}
  &+ R_{QCDC} \left\langle \hat a_{LL}^{QCDC} / D \right\rangle A_1\\
  &+ \sum_{f,f'=u,d,s,\bar u, \bar d, \bar s, g} R_{ff'} \left\langle \hat a_{LL}^{ff'} \left( \frac{\Delta f}{f} \right)^d \left( \frac{\Delta f'}{f'}\right)^\gamma \right\rangle. 
  \label{eq:aparoddecomp}
\end{align}
Note that we have neglected the small contributions of the leading process $\gamma^* q \rightarrow q$ and of the low $p_T$ scattering events because there is no hard scale allowing a perturbative treatment of these subprocesses (low transverse momentum, $Q^2<1$ GeV$^2$ events). 
As can be seen on Fig.~\ref{fig:rsubproc}, the fraction of photon-gluon fusion events is $R_{PGF}=0.31$. The analyzing power $\hat a_{LL}^{PGF}$ is the helicity asymmetry of the $\mu g \rightarrow \mu ' q \bar q$ scattering cross-section, $\hat a_{LL}^{PGF} \equiv d\Delta \sigma_{PGF}^{\mu g} / d\sigma_{PGF}^{\mu g}$. It is calculated from the kinematic variables of the partonic reaction for each PGF event in the high $p_T$ Monte-Carlo sample. Averaging over the PGF events, we obtain $\left\langle \hat a_{LL}^{PGF}/D \right\rangle = -0.933$. The contribution of the PGF process to the high $p_T$ asymmetry is thus $-0.292 \times \frac{\Delta G}{G}$. The contribution of the QCD Compton events is calculated in the same way to be 0.0063, using a fit on the world data for the virtual-photon deuteron asymmetry $A_1^d$.

Resolved photon subprocesses involve either a quark or a gluon from the nucleon. In the latter case, they are sensitive to the gluon polarization $\Delta G/G$, and contribute to the signal. The analyzing powers $\hat a_{LL}^{ff'}$ are calculated in pQCD at leading order~\cite{Bourrely:1987gp}, and are positive for all relevant channels. The polarizations $(\Delta f/f)^d$ of the $u$, $d$ and $s$ quarks in the  deuteron are calculated using the unpolarized parton distribution functions from GRV98, and the polarized parton distribution functions from GRSV2000~\cite{Gluck:2000dy}, at leading order. The polarizations of quarks and gluons in the virtual photon $(\Delta f/f)^{\gamma}$ are unknown as the polarized PDFs of the virtual photon  have not yet been measured. Nevertheless, theoretical considerations provide a minimum and a maximum value for each $\Delta f^\gamma$~\cite{Gluck:2001rn}, called the {\em minimal} and {\em maximal scenarios}. The total contribution of the resolved photon processes to the high $p_T$ asymmetry, which ranges between $0.000 + 0.012 \times \Delta G/G$ and $0.002 + 0.078 \times \Delta G/G$, will be taken into account in the systematic error on $\Delta G/G$. 

The tuning of the PYTHIA parameters relevant to this analysis is an important source of systematic errors. This error was estimated by scanning these parameters independently over a range where the  agreement between the simulation and real data remains reasonable. This resulted in several values for $\Delta G/G$, all based on the same high $p_T$ asymmetry, Eq.~(\ref{eq:measaparodall}). The value of $\Delta G/G$ appears to depend predominantly on the width of the intrinsic transverse momentum distribution for the partons in the photon. 
For instance, varying this parameter between 0.1 and 1~GeV/c results in a 30\% variation of the fraction of photon-gluon fusion $R_{PGF}$. NLO effects seem to be small: varying the scale and (de)activating parton showers does not affect the result.

Using Eq.~(\ref{eq:aparoddecomp}) to extract the gluon polarization from the high $p_T$ asymmetry, Eq.~(\ref{eq:measaparodall}), we finally get:
\begin{equation}
\frac{\Delta G}{G}(x_{g} = 0.095, \mu^2=3\ \operatorname{GeV}^2) = 0.024 \pm 0.089 (stat.) \pm 0.057 (syst.).
\label{eq:dgog20023quadsys}
\end{equation}
This result was compared to the recent distributions of $\Delta G(x)/G(x)$ from AAC~\cite{Hirai:2003pm} and LSS~\cite{Leader:2001kh}. These two distributions, which strongly differ in shape, are almost equal at $x_g = 0.095$ and are compatible with our result within $1.5~\sigma$. The first moments of $\Delta G$ are equal to $0.8 \pm 0.56$ and $1.1 \pm 0.52$ for the AAC and LSS fits, respectively. The fraction of PGF can also be enhanced by selecting events with charmed hadrons ($D^0$ and $D^*$), instead of high $p_T$ hadrons. The results obtained with this method, which is much less model-dependent but suffers from low statistics, will be produced soon.

\bibliography{cbernet_dis05proc}

\end{document}